# THE ENVIRONMENT EFFECT ON OPERATION OF IN-VESSEL MIRRORS FOR PLASMA DIAGNOSTICS IN FUSION DEVICES

V.S.Voitsenya*, Ch.Gil[1], V.G.Konovalov, A.Litnovsky[2], M.Lipa[1], M.Rubel[3], A.Sagara[4], A.Sakasai[5], B.Schunke[1], T.Sugie[5], G. De Temmerman[6], K.Yu.Vukolov[7], S.N.Zvonkov[7], P.Wienhold[2]

*Institute of Plasma Physics, NSC KIPT, Akademichna St. 1, 61108 Kharkov, Ukraine;*
[1]*CEA Cadarache, 13108 St.-Paul-Lez-Durance, France;*
[2]*Institut für Plasmaphysik, Forschungszentrum Jülich, D-42425, Jülich, Germany;*
[3]*Royal Institute of Technology, Ass. EURATOM-VR, S-100 44 Stockholm, Sweden;*
[4]*National Institute for Fusion Science, Oroshi-cho, Toki-shi, Gifu-ken 509-5292, Japan;*
[5]*Naka Fusion Research Establishment 801-1, Mukoyama, Naka-machi, Naka-gun, Ibaraki, 311-0193, Japan;*
[6]*Institut für Physik, University of Basel, Switzerland;*
[7]*RRC "Kurchatov Institute", Moscow, Russia*

*Corresponding author. E-mail: voitseny@ipp.kharkov.ua

First mirrors will be the plasma facing components of optical diagnostic systems in ITER. Mirror surfaces will undergo modification caused by erosion and re-deposition processes [1,2]. As a consequence, the mirror performance may be changed and may deteriorate [3,4]. In the divertor region it may also be obscured by deposition [5-7]. The limited access to in-vessel components of ITER calls for testing the mirror materials in present day devices in order to gather information on the material damage and degradation of the mirror performance, i.e. reflectivity. A dedicated experimental programme, First Mirror Test (FMT), has been initiated at the JET tokamak within the framework Tritium Retention Studies (TRS).

**Introduction**

About half of all methods of plasma diagnostics in ITER will be based on analysis of characteristics of electromagnetic radiation in different parts of spectrum. Due to high level of neutron and gamma radiation, all components of optical schemes nearest to burning plasma must be metallic mirrors because, in contrast to characteristics of refractive optical components (lens, prisms), the optical properties of metals are not significantly affected by both these deeply penetrating radiation. In each optical scheme one mirror has to be the plasma facing component, and just this mirror (first mirror, FM) will be subjected to strongest plasma impact.

For nearest to plasma mirrors there will be two main reasons resulting in degradation of the optical properties. (i) The long-term bombardment of FM surface by charge exchange atoms (CXA) will result in the increasing surface roughness. (ii) Another reason for degradation of optical properties of in-vessel mirrors is deposition of material eroded from the components subjected to strongest plasma impact.

Up to recent time the simulation experiments on behavior of in-vessel mirrors were provided using small-scale stands [1], but since 2001 the investigations of this problem are included in the experimental program of several large scale fusion devices and some useful experience was gained. In this paper the main results obtained on LHD, T-10, Tore Supra and TEXTOR are analyzed and some details of future experiments on JET are described.

**1. LHD**

One of the first attempts to study the correlation between the mirror location and modification of mirror optical properties was made a few years ago for stainless steel (SS) mirror samples exposed inside the Large Helical Device (LHD) during the whole 3rd experimental campaign (July 1999-December 2000) [2]. During that experimental campaign there were about

10,000 working discharges and about 2300 hrs of glow discharge cleaning (in hydrogen or helium).

The main peculiarities of LHD operating regimes during this campaign are described in [3]. Here we note that the graphite tiles were installed in the divertor areas in such a way that the plasma of divertor flows did interact with graphite targets only.

Below the main results obtained by studying the optical properties and surface composition of mirror samples taken out of LHD vessel are presented.

Three SS mirrors with size 20x10x1 mm were mechanically polished, cleansed in an ultrasonic bath filled with acetone, and then the spectral reflectance at normal incidence was measured in the wavelength range 200-700 nm. The samples were installed inside the LHD vacuum vessel in locations indicated in Fig. 1 by numbers 1, 3, and 5: the sample #1 was positioned near the divertor region, #3 – close the plasma edge, and #5 – quite deeply in the diagnostic port in the same poloidal cross section and in the same plane as #1 (central plane). After samples were removed the reflectance was again measured in the same way, and the surface of samples were analyzed by several methods: Auger electron spectroscopy (AES), ion backscattering technique (RBS) using 1.5 MeV $He^+$ ion beam, scanning electron microscopy (SEM), profilometry, and ellipsometry at the wavelength 632.8 nm. These data are presented in detail in [4].

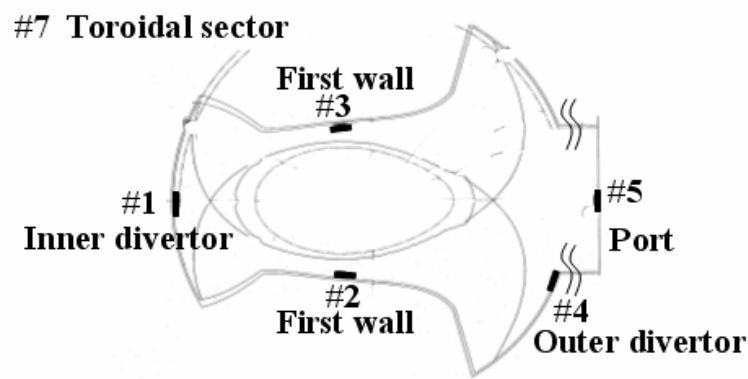

Fig. 1. The scheme of locations of SS mirror samples inside the LHD vessel.

The photos of Fig. 2 show how mirror samples looked just after finishing their exposure in LHD vacuum vessel. As a result of exposure in LHD the reflectance of all samples has changed from the identical initial level, Fig. 3.

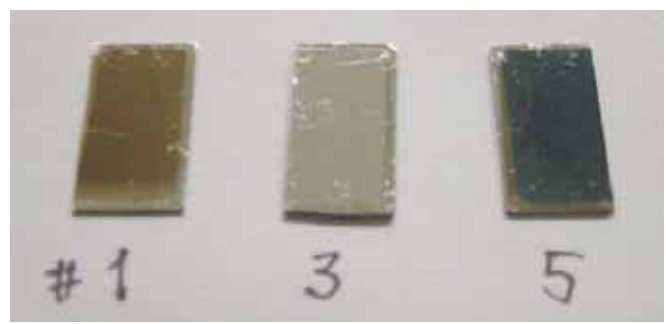

Fig. 2. The photos of samples after they were taken out of the LHD vacuum chamber.

The reflectance of mirrors #1 and #5 dropped strongly as a result of appearance of the contaminating films. From spectral dependence of reflectance one can conclude that the film on sample #5 is thicker than on the #1 sample. At the same time, the reflectance of sample #3 increased significantly and became close to what is usually observed when after rinsing in a

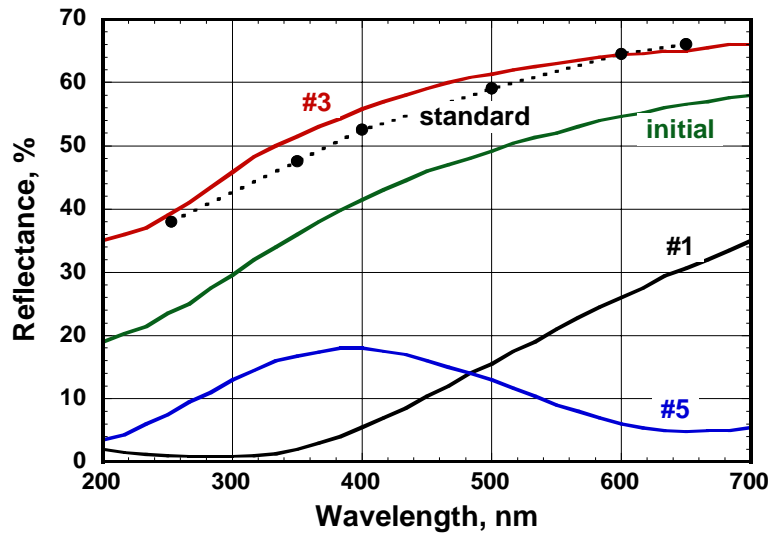

Fig. 3. Spectral reflectance of SS mirror samples before (marked as "initial") and after exposure inside the LHD vessel (curves marked as #1, #3, and #5). Solid points show the typical spectral reflectance of a SS mirror subjected to cleaning by low temperature deuterium plasma after polishing and washing in an ultrasonic acetone bath.

After rinsing in the ultrasonic bath the samples are cleaned by low temperature plasma of an ECH discharge in deuterium, as shown by solid points (the curve "standard"). This effect is connected with not full cleaning of all three mirrors before they were installed inside the LHD vessel from contamination by some organic film appeared due to rinsing samples in acetone. This initially existed film was fully cleaned from sample #3 due to plasma impact but became over coated by new films (to be immured) at samples #1 and #5. The high level of reflectance for mirror sample #3 measured after exposure samples in LHD was supported by very high quality of surface as follows from profilometry measurements (Fig. 4) and analysis of the SEM photos (not shown).

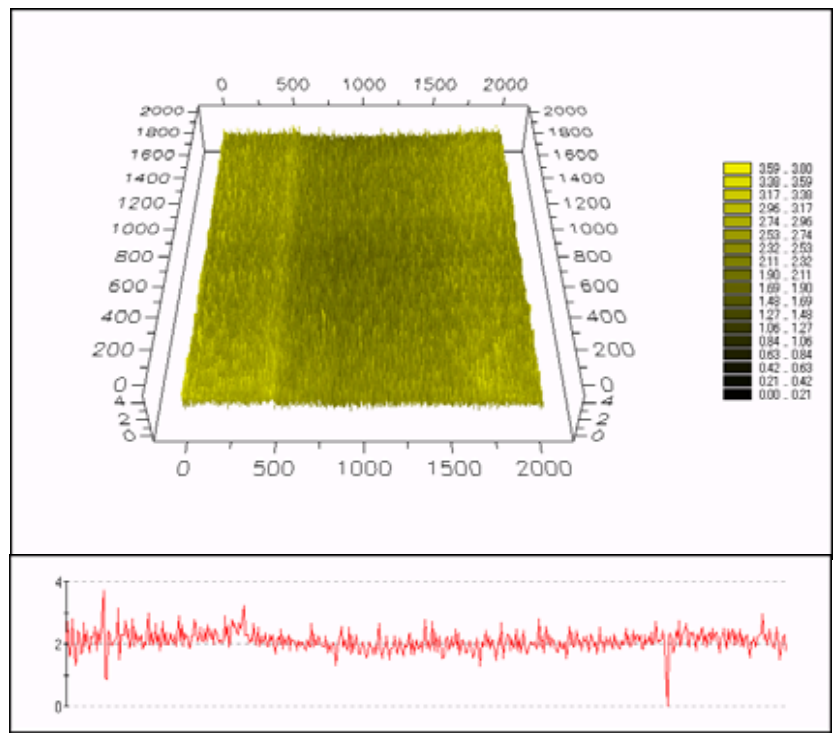

Fig. 4. Results of profilometry measurement of the surface morphology of the sample #3 after exposing in the LHD.

The optical indices of film on sample #1 measured by ellipsometry at the wavelength 632.8 nm were found to be n=2.5 and k=0.33 and the film thickness ~26 nm. For sample #5 similar data could not be obtained in similar way because of very low reflectance at the wavelength of measurement (Fig. 2). Therefore the initial thickness of this film was estimated by providing ellipsometrical measurement after the film was partly chemically eroded using deuterium plasma of ECH discharge [5] and amounted ~50 nm.

According to AES analysis, the deposited layer on the #1 sample consisted mainly of C (~40 atomic %) and Fe (~40 atomic %) but on the #5 sample – the only contaminant registered was carbon (~90 atomic %) [4]. The surface of the sample #3 was practically free of carbon however a quite small trace of heavier metal, possibly, copper was registered by RBS. As follows from profile measurements, the layer of ~0.15 μm was eroded from the surface of this mirror sample [4] due to plasma impact: CXA during working discharge and ion bombardment during wall conditioning.

According to AES measurements, the thickness of deposited layers on the collector probes disposed close to these mirror samples was ~50 nm and ~30 nm for samples #5 and #1 correspondingly [2], i.e., in good agreement with ellipsometry data.

## 1.1. Main results

Two mirror samples remote from the confined plasma became coated with carbon-based film. The characteristics and thickness of deposited films are very different: on the mirror placed in the divertor region the film was hard and included large portion of iron. Cleaning this mirror from contaminating film by low temperature deuterium plasma of an ECR discharge without acceleration of ions was not too effective. In contrast, the mirror placed deeply in the port was coated with soft C film containing only small percentage of hydrogen (~10%). It was much easily taken off by exposing to low energy ions of deuterium plasma [4].

The mirror placed in the nearest location to the hot plasma became to be cleaner than it was initially. The full erosion of this sample found by measurement of the surface profile did not exceed 0.15μm. Such small net erosion seems surprising if one takes into account the long time of glow discharge conditioning and large number of working discharges. Probably the balance between sputtering by CXA during working discharges and by plasma ions during conditioning, on the one side, and contaminant deposition, on the other side, was almost zero, only with minor preponderance of sputtering erosion over deposition.

## 2. T-10

Polished mirror samples of SS316 (10×10 mm$^2$ and 20×20 mm$^2$, 4 mm thick) were used for experiments. They were exposed in diagnostic port of limiter section of T-10 at different distances from plasma as shown in Fig. 5 [6].

Thin SS shield protected a part of each sample with ~0.1 mm gap between shield and mirror surface. Samples fixed in 1-3 positions were shutter-protected during vacuum vessel (VV) conditioning. The positions of mirrors 1-6 were frontal to plasma. Besides, in 2003 mirrors 8-9 were placed edgewise to plasma and one mirror located in frontal position, 7, was shadowed by sample 5.

The typical parameters of T-10 discharge are following: working gas – deuterium; plasma current – 200-400 kA during 1 s; toroidal magnetic field – 2.8 T; central electron temperature $T_e(0) \approx 1$ keV for the Ohmic phase and up to 2.5 keV at ECRH; line averaged electron density in the range $(1-6) \times 10^{19}$ m$^{-3}$; central ion temperature near 450-700 eV at both phases. Usual configuration with ring and movable limiters from graphite MPG-8 was used in 2002 but in 2003 the ring limiter was taken away and only movable limiter from graphite RGT-91 was used. Central plasma parameters were remained practically at the same level but the intensity of plasma wall interaction greatly increased.

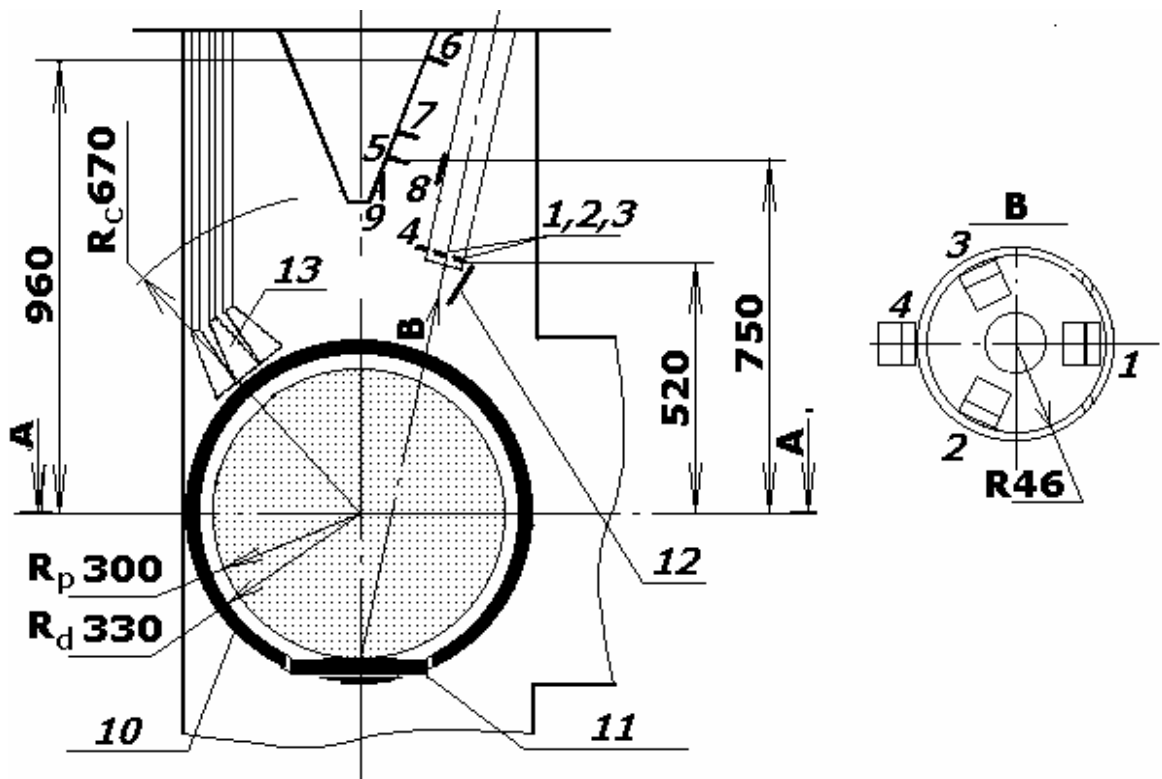

Fig. 5. The layout of experiments in T-10.
1-9 – mirrors (note, mirror #7 is in the shade of mirror #5), 10 - ring carbon based limiter, 11 - movable graphite limiter, 12 - shutter, 13 – ECRH antenna. Rc – radius of toroidal magnetic coils; Rp – plasma radius; Rd – radius of ring limiter.

Every experimental campaign started from vacuum vessel (VV) cleaning by heating up to 200°C, inductive $H_2$ or $D_2$ discharges and He or Ar glow discharges. During experiments these procedures of VV conditioning were used every night. The total duration of VV conditioning modes and plasma discharges are given in Table I.

Table I. The data on VV conditioning modes and plasma discharges

|      | Heating up to 200°C (hours) | Inductive discharges (hours) | Glow discharges (hours) | Plasma discharges (s) |
|------|-----------------------------|------------------------------|-------------------------|-----------------------|
| 2002 | 897                         | 35 ($H_2$) 270 ($D_2$)       | 86 (He)                 | 1620 ($D_2$)          |
| 2003 | 367                         | 24 ($H_2$) 79.5 ($D_2$)      | 19 (He) 2 (Ar)          | 212 ($D_2$)           |

The reflectance of mirrors before and after exposition and film thickness were measured by means of Lambda 35 (Perkin Elmer) spectro-photometer and with TENCOR Instruments profilometer, correspondingly. The surface morphology and microstructure of deposits were investigated with JEOL scanning electron microscope (SEM) and by X-ray diffraction. The Rutherford Back Scattering (RBS) in combination with the resonance elastic scattering were used for analysis of deposits composition. RBS analysis was carried out with 2 MeV protons backscattered from the sample at the angle of 170°. Hydrogen isotopes depth's profiles were measured with Elastic Recoil Detector Analyses (ERDA) by the $He^+$-ion beam with the energy of 1.9 MeV.

The deposited films have arisen on all samples after exposure in T-10 during both 2002 and 2003 runs. It means the predominance of deposition over erosion in the given experimental conditions. Results of 2002 experimental run were presented in [7,8]. The surface of mirrors looked shining and smooth and the color of deposits varied from golden-crimson to violet-green. The screened parts of mirrors practically did not changed by sight, Fig. 6. The deposits not only decrease the reflectance of mirrors but also distort the spectrum of reflected radiation because of interference phenomenon.

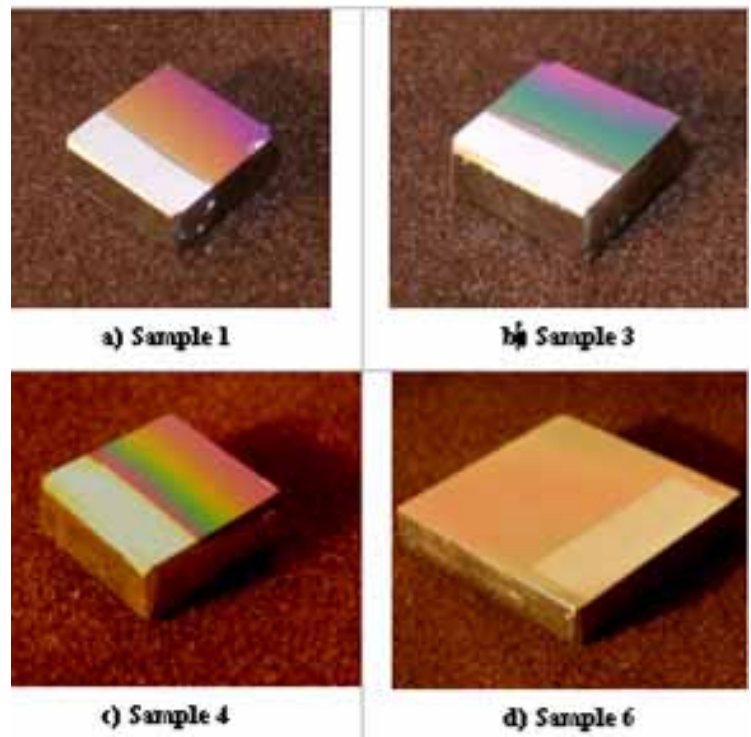

Fig. 6. Photos of samples after finishing the 2002 year campaign. Lighter parts of mirrors were closed during whole exposure time. The deposit has quite homogenous, fine-dispersed structure. On mirrors protected by shutter during conditioning it grew with the rate 0.2 nm/s in 2002 and 4 nm/s in 2003.

Reflectance of mirrors placed edgewise to plasma (8 and 9) changed only slightly and no reflectance change was observed for mirror in position 7.

The deposits have a quite homogenous, fine-dispersed structure with inclusions of single spherical particles up to 1-3 µm in size and flakes having dimensions up to 25 µm [8]. Their origination can be explained by brittle erosion of the limiter under plasma disruptions.

According to RBS data the deposits basically consist of carbon, hydrogen isotopes and of about 5-8% oxygen impurity. Deuterium prevails over hydrogen on the opened parts and its concentration is practically constant inside the film, in the range 25-55 %.

Hydrogen was found only in the near-surface layer down to the depth <80 nm. The thickness of deposits on screened parts of mirrors did not exceed 40-50 nm. Note that hydrogen, deuterium and carbon were uniformly distributed in the thin film (≈80 nm) on the mirror 9. Possibly this film rose predominantly during VV conditioning when hydrogen was used, similar to the deposits on screened parts of mirrors.

The film thickness, the integral concentrations of deuterium and D/C ratio in deposits on the opened parts of mirrors are presented in Table II.

Table II. Main parameters of deposited films (*samples protected by shutter at VV conditioning).

| Sample Position | Distance from plasma, cm | Thickness of films on opened part, μm | | D integral concentrations ($10^{21}$ m$^{-2}$) | | D/C | |
|---|---|---|---|---|---|---|---|
| | | 2002 | 2003 | 2002 | 2003 | 2002 | 2003 |
| 1* | 22 | 0.2 | 0.3 | - | 18 | – | 1.6 |
| 2* | 22 | 0.3 | 0.7 | 6.5 | 28 | 0.2 | 1.9 |
| 3* | 22 | 0.4 | 0.9 | - | - | – | – |
| 4 | 22 | 12 | 2 | - | 20 | – | 1.3 |
| 5 | 45 | 3.9 | 0.6 | - | 32 | – | 2.2 |
| 6 | 66 | 1.0 | 0.3 | 7.5 | 16 | 0.35 | 2 |
| 7 | 47 | - | no film | - | - | | - |
| 8 | 45 | - | 0.1 | - | - | | - |
| 9 | 40 | - | 0.1 | - | 1.5 | | 0.4 |

Note that D/C ratio was equal 0.2-0.35 in usual operation mode in 2002 but it increased about 10 times in 2003 probably due to intensification of the plasma-wall interaction. For the mirrors "protected" during VV conditioning the deposition rate was estimated from known duration of plasma discharges and the film thickness. The deposition rate was about 0.2 nm/s in 2002 and 4 nm/s in 2003. Importantly that only thin films were found out on the mirrors placed edgewise to plasma (8 and 9) and the film was practically absent on the surface of mirror 7 located in shadow of sample 5. It gives hope that in ITER conditions the secondary mirrors will not be covered with contaminating film.

The X-ray analysis showed two lines of CH-crystalline but no free carbon inside the deposits [8]. The 12 μm hydrocarbon film (sample 4, 2002) was transformed into a friable, black carbon based film in the course of annealing in vacuum step by step at fixed temperature within 10-20 minutes. Gas began to desorb at 450°C and finished at 600°C. During vacuum annealing the molecules with different combinations of deuterium and hydrogen with carbon and oxygen were detected. So, it may be supposed that the deposits basically contain different $C_nD_m$ compounds.

Recently the ablation of the deposits was investigated under impact of pulsed excimer laser ($\lambda$=308 nm, F=10 Hz, $\tau$=20 ns) in high vacuum conditions [8]. The laser beam was focused on a surface of mirror in a rectangular spot with size 6×2 mm$^2$ at the energy density 0.05-0.15 J/cm$^2$. The total removal of deposits was observed after approximately 30 laser pulses at the energy density higher than 0.12 J/cm$^2$. When the energy density was a little less (about 0.1 J/cm$^2$) the deposit was not removed and only the surface flaking was observed. The SEM analysis of laser track showed that deposits have a layered structure.

### 2.1. Main results
Deposition prevails over erosion for the mirrors located in upper limiter port of T-10. Deposited films not only reduce the intensity of the reflected light, but also strongly change its spectra. Deposition rate and film composition depend on distance and orientation of mirrors relatively plasma. Results of 2003 campaign show that deposition rate and D/C ratio can be very high in particular modes of tokamak operation. Mirrors shaded by obstacles from the plasma direct vision can be possibly protected from the fast growing of the deposit.

### 3. Tore Supra
Two pairs of Mo, SS and Cu mirror samples, prepared by Kharkov IPP and Kurchatov Institute, were installed in the vacuum vessel of the Tore Supra tokamak in February 2003. The molybdenum samples were fabricated from a single crystal (sc) Mo(110) with divergence of surface orientation ~30 minutes. The Mo samples were mechanically cut from the rod then mechanically treated and finally polished in a plant "Luch" in Podol'sk (Moscow region). The stainless steel samples were fabricated from a polycrystalline material. This steel was developed in the USSR as an analogue of the AISI 316 steel. The official mark of this steel is 04Cr16Ni11Mo3Ti, which means ~0.04% carbon, 16 % Cr, 11% Ni, 3% Mo and some small amount of Ti. The base is iron. The polycrystalline copper samples were made of oxygen free copper and subjected to diamond turning, as high quality mechanical polishing of such soft material seems to be difficult.

The mirror samples of 22 mm diameter and 4 mm thickness have been installed on the high field side of the TS plasma vessel. They were located at theta = 7.5° and 15° out of the equatorial plane (Fig. 7) and positioned roughly 140 mm from the LCFS. The reflecting surface of all mirrors were oriented parallel (not inclined) to the toroidal field direction. The mirror assembly allowed contact cooling via the first wall steel panels. 2D- thermohydraulic finite element calculations of the mirror sample assembly showed temperatures not exceeding ~150°C for a typical "Gigajoule" operation scenario (~1 MW radiated plasma power loss during 36 s of pulse length).

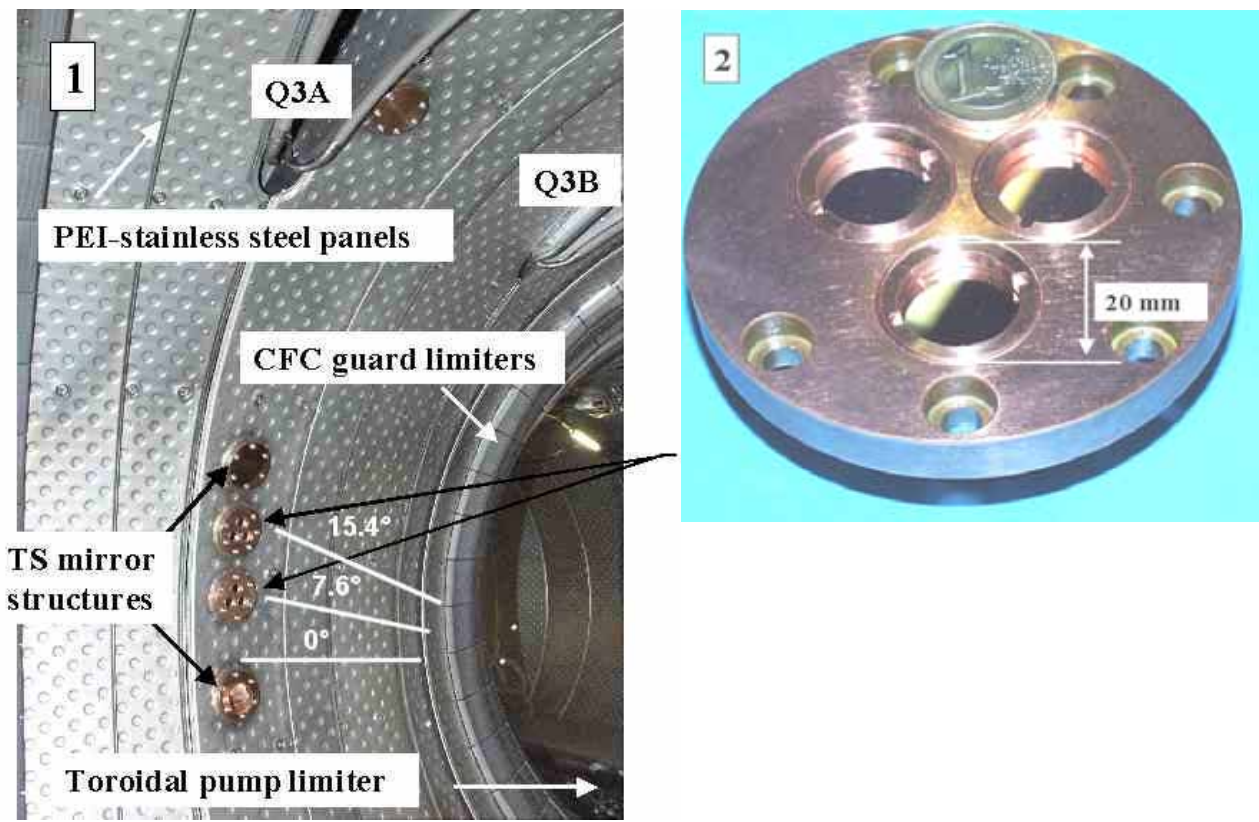

Fig. 7. Mo, SS and Cu mirror samples assembled in CuCrZr ring holders and fixed on Tore Supra vessel wall at high field side.

### 3.1. Main result [9]

Total and diffusive reflectance values were measured before and after exposure in the Tore Supra tokamak by means of a spectrophotometer ($\lambda$=250-2500 nm) and significant change of reflectance was observed for all samples, but especially for Cu, as Fig. 8 shows. From measurements by con-focal microscopy the surface roughness of the Cu mirror samples after exposure increased by factor 7 and a net erosion profile depth amounted ~2.8 $\mu$m, while the surface roughness of the sc Mo samples did not change with a net erosion profile depth only ~0.12 $\mu$m.

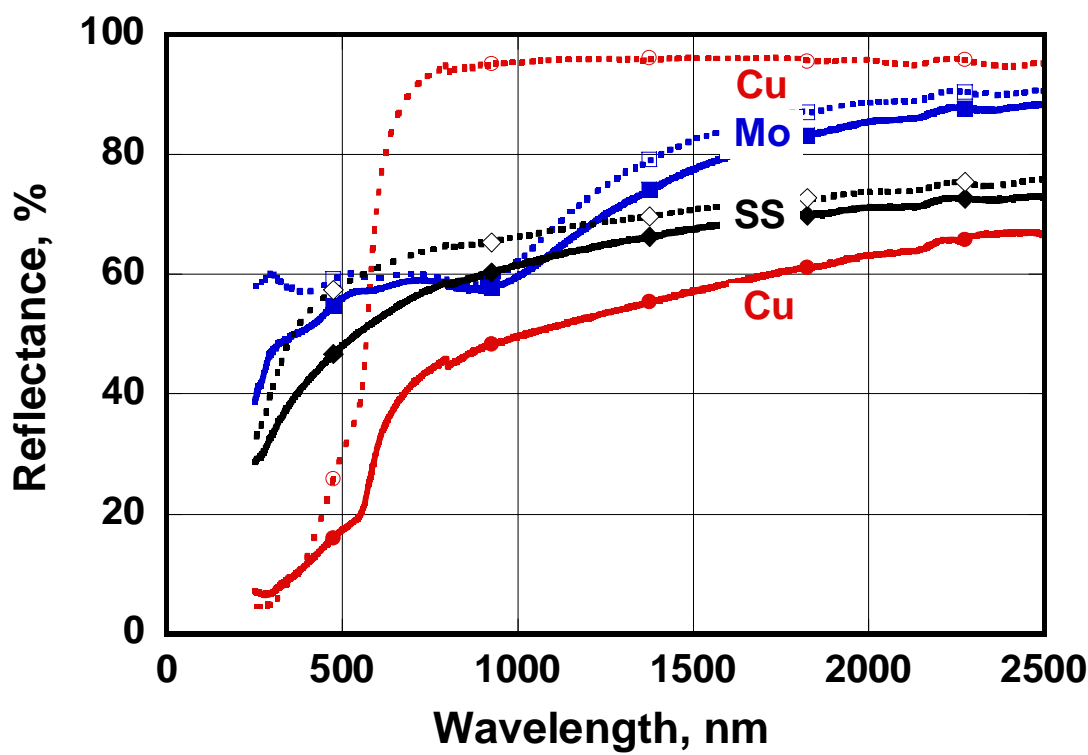

Fig. 8. Reflectance of mirror samples before (doted lines) and after (solid lines) exposure inside Tore Supra.

It was found that all exposed mirror samples became not only eroded but also coated with some deposited layers what is the main reason for change of the total reflectance.

Independent measurements at IR wavelength (118.6μm) showed insignificant variation for all mirror samples.

Ex-vessel glow discharge experiments in progress on virgin mirror samples show that long lasting conditioning procedures can contribute to additional important material erosion.

## 4. TEXTOR

Polycrystalline mirrors from molybdenum were placed in the SOL plasma of TEXTOR [10]. The photograph in Fig. 9 shows the experimental set-up and the appearance of mirrors after exposure. One large mirror (115mm x 74mm) was fixed inclined at 20° to the toroidal fluxes. The mirror was partly protected by an aluminium bar which created a shadow. Outside the shadow some deposition becomes visible. One small mirror was placed for comparison at the rear side of the holder perpendicular to the toroidal direction. All mirrors were polished mechanically and had optically been pre-characterized.

In the first experiment the edge of the mirror closest to the plasma was kept at a distance of 8 mm from the LCFS (plasma radius 46 cm) which is an erosion dominated zone in TEXTOR. It was exposed to 30 neutral beam heated (increasing from zero up to 1.3 MW) plasma pulses (line averaged density $n_e = 1.5 \times 10^{19}$ m$^{-3}$) with a total exposure time of 197 s. These conditions caused strong temperature rises (>1000°C for ≈ 3s) of the edge of the small mirror. The temperature of the large mirror measured by a pyrometer rose to 400-750 °C during the discharges. For the second experiment the closest edge of the mirrors to the plasma were placed at LCFS+25mm to cover primarily the deposition dominated zone. They were exposed to 58 (partially neutral beam heated) discharges with varying plasma density ($n_e = 1.5$ to $4.5 \times 10^{19}$ m$^{-3}$) and a total exposure time of 312 s. The bulk temperature achieved 150°C, local temperature rises have not been observed. Carbon deposition up to thicknesses of more than 200 nm took place except in shadowed regions and at the edges. After dismounting the surface the deposits were investigated by ion-beam analyses.

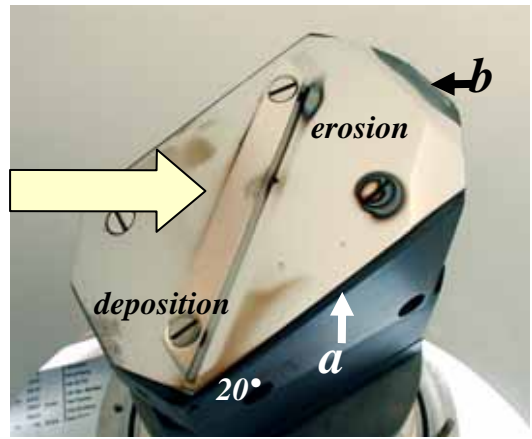

Fig. 9. Photograph of the mirror holder after exposure in the erosion dominated zone. The large mirror (a) is mounted 20° inclined on the graphite carrier, the small mirror (b) on the rear side perpendicular to the toroidal direction.

The set-up of a periscope-like system on TEXTOR simulates the ITER diagnostic ports transmitting the plasma radiation to further distant sensors. Two plane mirrors were installed in tubes facing each other at a distance of 275 mm. The mirrors are oriented 45° with respect to the line of sight. The orifice (35 mm diameter) is oriented perpendicular to the toroidal direction.

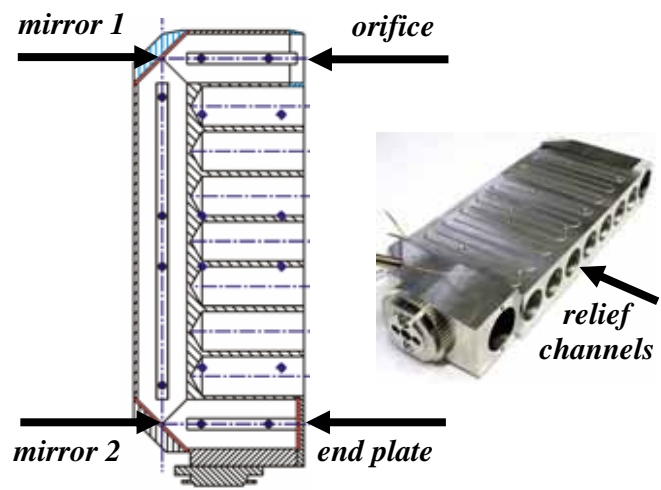

Fig. 10. Periscope system with mirror 1 (closest to the plasma), mirror 2 and the end plate. The tubes (35mm) are wrapped inside with exchangeable steel foils fixed at the wall by bars. The photograph shows the stainless steel body without end plate. Relief channels reduce weight.

The second end was closed with a polished end plate as is shown in Fig. 10. The SOL fluxes hit the first mirror under 45°. Carbon transport inside and deposition can be investigated after dismounting of the system. The deposition patterns on exchangeable steel foils wrapped inside the tubes complete this information. The stainless steel body can be heated up to about 500°C. Mirrors made from TZM (99% Mo, 0.1% Zr, 0.5% Ti) were used for the pilot experiment. The edge closest to the plasma was kept at 25 mm from the LCFS and the system exposed to 89 (partly neutral beam heated) plasma pulses ($n_e$ = 1.5 to $4.5 \times 10^{19}$ m$^{-3}$) with a total plasma exposure time of 1047 s. The temperature was 150 °C.

Total and diffuse reflectivities were measured under almost normal incidence (3° 20') before and after exposure by means of a spectrophotometer (wavelength range λ=250-2500 nm) with a space resolution of ≈10 mm. Fig.11 shows examples for the wavelength dependence of the total reflectivity $R_{tot}$ found on the two large mirrors. The central curves (1) represent the total reflectivity measured before and after plasma exposure at a location protected by the aluminum bar. The lines stay below the values given for polycrystalline molybdenum in literature [11]. As soon as deposition begins (curve 2 in Fig.11a) the reflectivity decreases by absorption in the film (thickness ≈30 nm). The drop is most pronounced in the UV region and levels off in the IR. In addition to absorption, the reflectivity can be distorted significantly due to destructive and constructive interference in thicker deposits. The example (curve 2, Fig. 11b) was measured on the mirror plate exposed deeper into the SOL (LCFS+25mm) where carbon deposition dominates and a film up to about 220 nm thickness was formed.

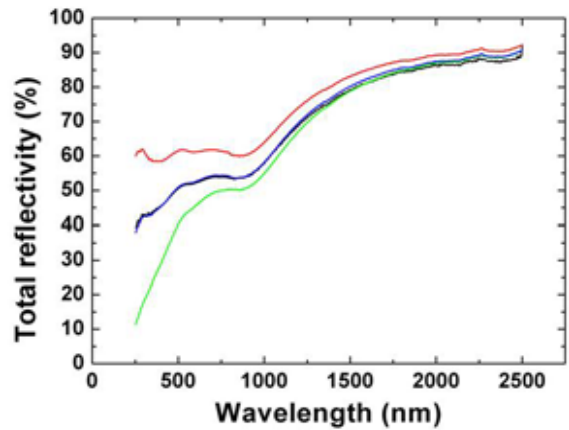

a)

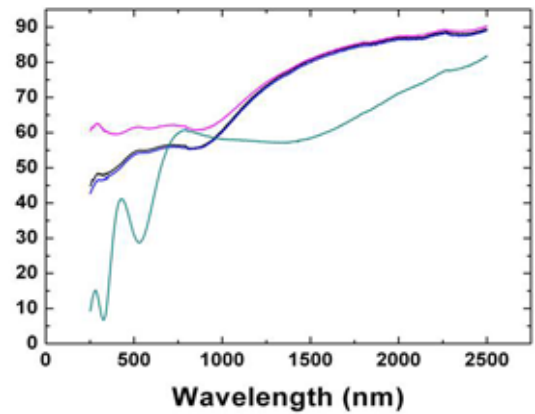

b)

Fig. 11. Total reflectivity $R_{tot}$ depending on wavelength measured on the molybdenum mirrors before and after exposure in a protected area (1) and after exposure in the deposition (2) and erosion (3) area. The mirrors (a) were exposed for 197 s in the erosion dominated zone (edge at LCFS+8mm) and (b) and for 312 s in the deposition dominated zone (edge at LCFS+25mm).

Surprisingly, an increase of $R_{tot}$ was measured on surface areas which suffered from net erosion (curves 3). The increases are similar, independent whether the target mirror was exposed near the plasma edge (Fig. 11a) or deeper into the SOL (Fig. 11b). This increase is most pronounced ($\Delta R_{tot}$ ≈10-20%) for short wavelengths and almost vanishes (1-3%) beyond about 1500 nm.

The local dependence of the reflectivity was measured at about N=50 different spots by shifting the mirror plates in the two directions (x and y) parallel to the sides. Before exposure the reflectivity scattered a little, i.e. the differences $\Delta R = R_i(\lambda) - R(\lambda)_{av}$ between one local measurement (index i) and the average out of the N measurements was less than about ±3 % in the visible and the IR. For shorter wavelengths the scatter was about twice as much. After exposure the local changes were much larger. Fig. 12 shows the N differences $\Delta R_{tot} = R_{after} - R_{before}$ measured on the mirror exposed at LCFS+8mm. The differences are highest for $\lambda=250$ nm (a) and achieve more than +12% (dark blue) in the eroded part of the mirror, while reductions of -30% (deep red) occur in areas covered with deposits. In protected or shadowed areas $\Delta R_{tot}$ is in the scatter range $\Delta R$ observed before exposure (blue green). For increasing wave lengths (b) the changes are less pronounced. The intensity of the diffuse reflected light has also been measured at the identical locations, but contributes little since the differences do not exceed about ±2% on this mirror plate.

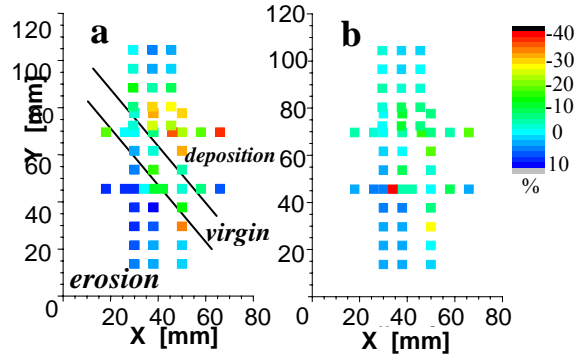

Fig. 12. Differences $\Delta R_{tot} = R_{after} - R_{before}$ given in colours at different surface locations for $\lambda=250$nm (a) and 500nm (b). The co-ordinates x and y cover the mirror dimension (74mm x 115mm).

An increase of $R_{tot}$ in the erosion dominated areas is expected, because the mirrors used in the experiments were not especially cleaned before installation in TEXTOR. The bombardment with plasma ions leads to a reduction of the surface impurity content and shifts the reflectivity up to the standard values given in literature for the well conditioned surfaces [9]. XPS-measurements [12] on layers formed by Mo with increasing contents of C or O show decreasing reflectivity and lead to the same conclusion. The surface erosion which is estimated to be less than 1µm [13] is mainly due to energetic deuterium and carbon ions. Part of the carbon remains intermixed in the near surface layer [14]. This is confirmed by SIMS depth profiling in the affected areas. But it has to be shown whether the presence of carbon in the interaction depth of ≈5 nm can influence the reflectivity.

The decrease of the total reflectivity in deposition dominated areas is due to absorption and by destructive interference if the deposit is thick enough. The example given (curve 3 in fig. 11b) roughly fits the condition $i*\lambda/4 = n*x$ (i=1, 2, 3..) for destructive or constructive interference using a film thickness of x=220 nm and a refractive constant n=1.65. The term n*x deviates by about +10% in the IR and by –10% in the UV region. This is likely due to the varying thickness within spot size of the measurement. But diffraction of light cannot be excluded. Combining the results from SIMS depth determination and the deuterium analysis with NRA yields a ratio D/C ≈ 0.3. The optical constants (n=1.65, k=0.02 for $\lambda=632$nm) were measured in the deposition zone of the molybdenum mirror. For ITER, a deposition of carbon on optical components is difficult to predict quantitatively, but it is possible because of the chemical erosion and transport of the hydrocarbon radicals over long distances.

Mirror #1 exposed in the periscope for 1047 s showed a radial decaying deposit with a maximum thickness of about 300 nm, Fig. 13, as follows from electron probe micro-analysis (EPMA). The thickness corresponds to a deposition rate of about 0.4 nm/s (inclination angle 45°). Although the deposition rate on mirror 1 is high, no detectable deposition was found on the following mirrors and walls.

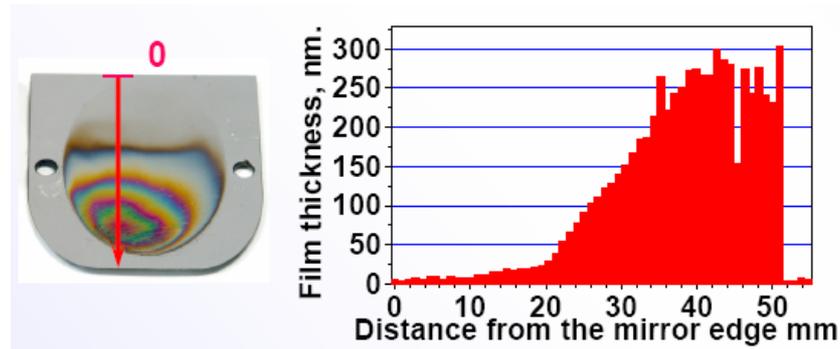

Fig. 13. The thickness of carbon-based film deposited on the mirror #1 of the periscope system.

The deposition rate of ≈ 0.4 nm/s on mirror #1 (inclination angle 45°) in the periscope is less than found on the 20° inclined mirror (0.7 nm/s) at the comparable radial position. This might partly be due to the fact that roughly one half of the carbon entering the orifice becomes deposited on the inner tube wall, visible there as a transparent deposit. Part of the carbon will likely be re-eroded from mirror #1. It is surprising therefore that no further deposition could be detected on the following walls and mirrors. One reason could be that the exposure time of 1047s is too short to grow a detectable layer. Another reason might be the pressure built up in the closed system which prevents the hydrocarbon radicals to penetrate deeper. Enhanced re-erosion of freshly formed deposits [15] may also play a role.

**4.1. Main conclusions**
The experiments show that the reflectivity of metal mirrors drastically changes after the exposure to the SOL plasma in TEXTOR. Increases to values close to standard values are observed in erosion dominated areas, as well as decreases of the reflectivity due to carbon deposition. Both effects are a concern for the mirrors in the ITER diagnostic ports and periscopes and can occur side by side on a mirror. It is obvious that the decrease of the reflectivity is due to absorption and destructive interference in the deposit, but for the observed increase it seems that the conditioning of the surface plays an important role. All changes of the reflectivity are the most pronounced in the near UV region, the IR region is almost not affected.

No detectable amounts of carbon were found on surfaces inside the periscope system beyond mirror #1. This behavior could be favorable for ITER, but the reasons for it remained unclear. These experiments are therefore a first step only to the understanding and developing of a model which can predict the ITER mirror performance.

**5. JT-60U**
In JT-60U, a tungsten plane mirror was installed near the divertor to measure impurity lines emitted in the divertor with a vacuum ultra violet (VUV) spectrometer.

In the case of the open divertor from 1991 to 1996, mitigation structure such as baffles against particle deposition on the surface had not existed. Hence the reflectivity of the mirror was significantly reduced with contamination which was mainly made of hydrocarbon coating.

After the modification to the closed divertor (so called the W-shaped divertor) in 1997, a tungsten plane mirror was installed to view the divertor plasma as shown in Fig. 14. The tungsten mirror, which was fixed by a box type SS support, was electrically insulated from the vacuum vessel with ceramics plates around the mirror in order to reduce particle deposition, particularly during the Glow Discharge Cleaning. No shutter is installed in front of the mirror. The mirror is located at the distance of 240 mm from the plasma. A molybdenum rectangular tube (25 mm in outer width, 1.5 mm in thickness and 140 mm in length) is used to reduce the viewing solid angle to 0.01 str. The angle of the sight line is 72˚ to the separatrix and 82˚ to the magnetic field line. The temperature of the mirror is around 300 °C since the vacuum vessel of JT-60U is 300 °C during the operation. The normal pulse length of JT-60U is about 15 sec. Under the condition, the reflectivity degradation of the mirror is not serious. VUV spectral lines can be observed with good S/N ratios after ~ 4000 shots of JT-60U.

From the experience of the tungsten mirror to observe the divertor plasma in JT-60U, the followings are suggested.

1) Something like baffles or long tube to minimize the viewing solid angle with matching the measurement should be used to avoid appearance of contamination (hydrocarbon coating). The use of the long tube is very effective to block off contaminants.

2) During the Glow Discharge Cleaning, the use of electric insulators from the vacuum vessel is strongly recommended to reduce introduction of Glow Plasma to the surface of the mirror. An electrical floating potential of the mirror by insulation seems to be better than the potential of the vacuum vessel to avoid contamination.

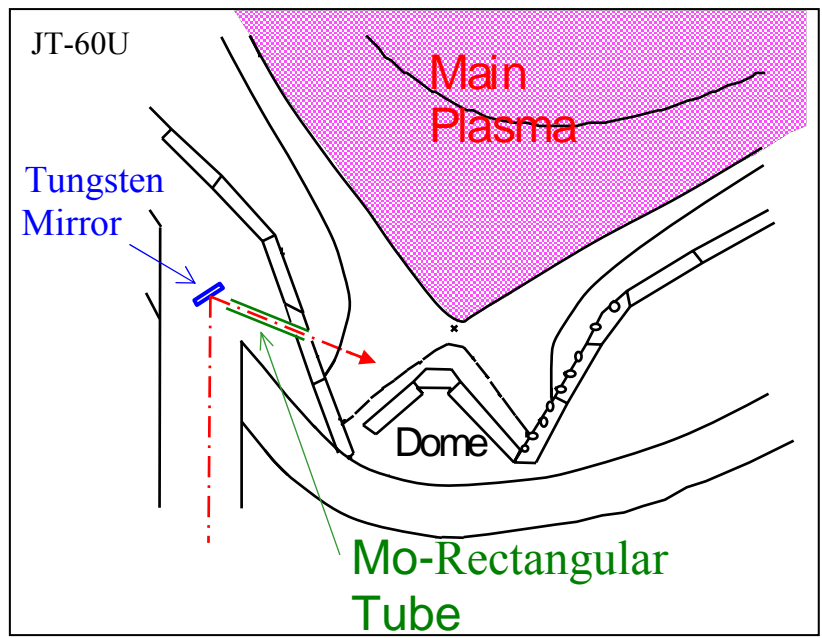

Fig.14. Disposition of W mirror for measurement of VUV and UV radiation emanating by the divertor plasma in JT-60U.

## 6. Conclusion

1. Deposition of contaminants on the in-vessel mirrors is probably impossible to avoid. The question is only in the rate of deposition or in the relation of deposition and sputtering rates.

2. Only those first mirrors maximally open to the hot plasma and fixed at the not far distance from the last closed magnetic surface can remain free from deposition of contaminating film due to possibility that the CXA sputtering rate predominates over the rate of film deposition (**LHD, Tore Supra**). However, the absence of contaminant layer on these mirror samples does not mean the entire lack of deposition: in reality both processes, sputtering and deposition, are probably simultaneously taking place but with some predominance of sputtering.

For such mirrors the guarantee of their long-term capacity for work is fabrication them from the single crystalline material. However, more data are necessary from these and other fusion devices, as **in T-10** all plasma facing mirror samples became coated with C films.

3. Secondary mirrors, shaded from the direct plasma view by obstacles, possibly can be protected from the growing the deposit (**T-10**, **TEXTOR**), but this statement must be checked in experiments on other large scale fusion devices.

4. The long tube minimizing the viewing solid angle with matching the measurement can probably be used to avoid appearance (or decrease of the rate of appearance) of contamination (hydrocarbon coating).
5. It is evident that the scale of experiments relating to investigations of environment effect on behavior of in-vessel mirrors in fusion devices under operation must be enlarged.

## 7. Future experiment at JET
### 7.1. Introduction

The environment conditions for mirrors in divertor region of ITER will be very different from those in the main chamber. The energy of the plasma particles will be <10 eV, i.e., below the threshold energy for physical sputtering of any material. Thus, for mirrors in divertor region (DFM) the sputtering will probably not play any significant role. Instead, the deposition of eroded material will be the principal process. If the carbon-based tiles are used in the divertor, the carbon will be the main impurity component in the divertor plasma. The carbon atoms will be in different state, and after hitting the surface they will combine with D-T atoms creating the radicals and volatile molecules, which can move freely through the slits into the region where divertor mirrors have to be situated. The carbon layer thickness can increase with time leading to the deterioration of mirror reflectance unless an effective method to suppress the deposition is developed. In addition, dust particles can be deposited on the DFM surface leading to a decrease of the reflectance and increase the scattering.

To clear up to a certain extent the role of both these processes, a dedicated experimental programme, First Mirror Test (FMT), has been initiated at the JET tokamak within the framework Tritium Retention Studies (TRS).

### 7.2. Scientific Rationale and the Experimental Programme at JET

The choice of JET for doing the FMT is related to several unique features of this machine:
- (a) the largest divertor tokamak with an ITER relevant configuration,
- (b) plasma pulses of 20 s,
- (c) beryllium environment,
- (d) comprehensive overview of erosion and deposition in concurrent deposition monitoring activities within the TRS Project.

The programme aims at the study of morphology changes occurring on surfaces of selected mirror materials: polycrystalline molybdenum and stainless steel. The inclusion of the FMT project in the framework of the TRS project ensures a comprehensive insight into erosion and re-deposition processes using a set of dedicated tools such as a quartz micro-balance (QMB) [16], indexable collector and other deposition monitors. Tested mirror samples will be installed in several places in the divertor and on the main chamber wall in the vicinity of these devices.

### 7.3. Mirror Materials

Following contacts with the ITER Team and the ITPA Diagnostic Expert Group two materials (out of 13 considered) were selected:
(a) Polycrystalline molybdenum
(b) Stainless steel 316L

The idea of testing mirrors coated with thin evaporated films was eliminated, both due to the cost involved and potentially greater risk of damage to surfaces during the exposure. Secondly, the test with only two different materials was dictated by the limited space for mirror carriers in the divertor and on the main chamber wall. The choice of solid mirror materials was discussed and agreed with the diagnostics experts and the ITER Team.

### 7.4. Design of Mirrors and Mirror Carriers

Fig. 15 exemplifies a design of a stainless steel cassette with several channels in which the mirrors are mounted. The number of channels in a cassette, three or five, is dependent on the available space in various locations of the machine: three positions in the divertor (two cassettes in the inner, two in the outer two in the base) and two on the main chamber wall.

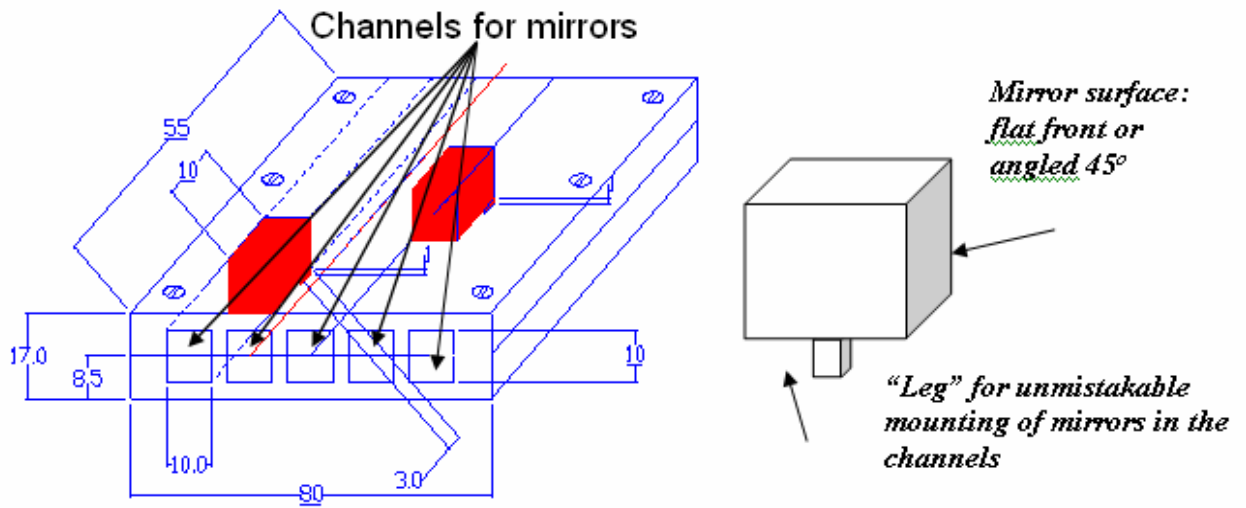

Fig. 15. A schematic view of a cassette and a mirror block for the test in JET.

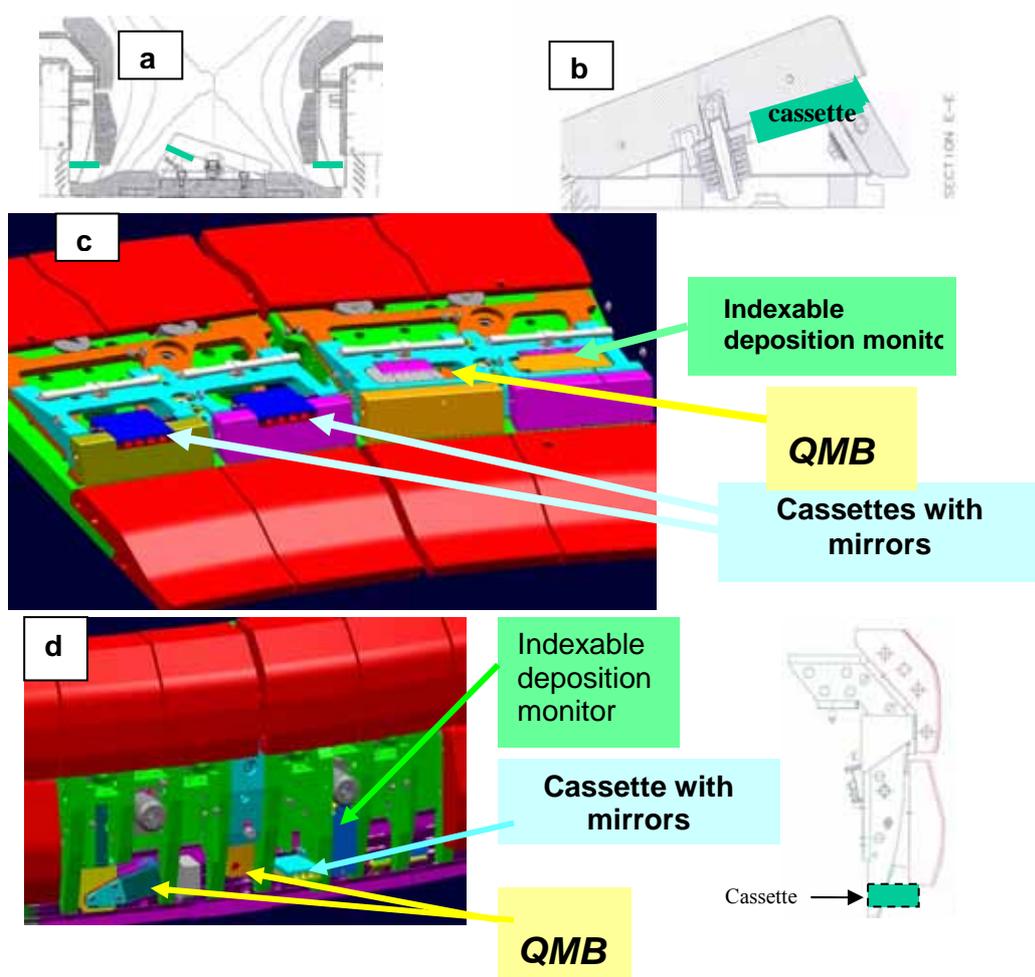

Fig. 16. (a) Divertor cross-section with a mirror cassettes (green boxes) in the inner, the outer carrier and below the load bearing plate (LBP made of carbon fibre composite); (b) base divertor module with a cassette; (c) location of several devices for monitoring the deposition mounted in the base divertor modules [Load Bearing Plate is removed]; (d) location of tested mirror and other deposition monitors in the inner divertor.

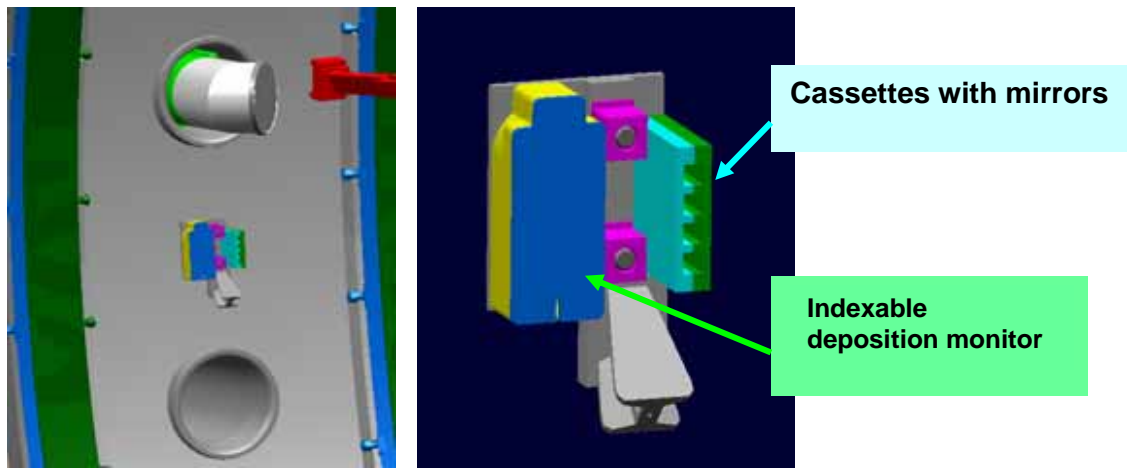

Fig. 17. Tested mirror and other deposition monitors in the main chamber wall mounted a clip (wall bracket) compatible with remote handling.

The cassettes are of a "pan-pipes" shape and they are composed of two detachable plates. The construction enables studies of deposition both inside the pan-pipes channels and on the mirror surfaces. Moreover, the exposure of mirror surfaces at various inclination angles and aspect ratio with respect to plasma is possible. Effects of sample temperature on deposition in the divertor may be studied by comparison with the QMB data [16]. Figures 15 a-c show the location of mirrors and other deposition monitors in the divertor base and the inner carrier. The arrangement of devices in the outer divertor is similar to that in the inner leg (see Fig. 15 d). There are also cassettes placed on the main chamber wall (Fig. 16) protected by magnetic shutters which open when the magnetic field is on. This ensures avoiding the deposition of beryllium during its evaporation on the JET wall. Installation of all the deposition monitoring devices in the JET vessel is compatible with remote handling [17]. The construction phase and commissioning of the diagnostic tools for TRS (including FMT) is to be ready for installation in 2004. The experiment itself will be carried out in 2005 until the next opening of the JET vessel.